**The problem of Gauge invariance in Dirac field theory**

By

Dan Solomon

Rauland-Borg Corporation
3450 W. Oakton
Skokie, IL 60076
USA

Phone : 847-324-8337
Email : dan.solomon@rauland.com

June 5, 2003

**Abstract**

Quantum field theory is assumed to be gauge invariant. However it is well known that when certain quantities are calculated using perturbation theory the results are not gauge invariant. The non-gauge invariant terms have to be removed in order to obtain a physically correct result. In this paper we will examine this problem and determine why a theory that is supposed to be gauge invariant produces non-gauge invariant results.

## **I. Introduction**

Quantum field theory is assumed to be gauge invariant [1][2]. A change in the gauge is a change in the electric potential that does not produce a change in the electromagnetic field. The electromagnetic field is given by,

$$\vec{E} = -\left(\frac{\partial \vec{A}}{\partial t} + \vec{\nabla} A_0\right); \quad \vec{B} = \vec{\nabla} \times \vec{A} \tag{1.1}$$

where $\vec{E}$ is the electric field, $\vec{B}$ is the magnetic field, and $\left(A_0, \vec{A}\right)$ is the electric potential. A change in the electric potential that does not produce a change the electromagnetic field is given by,

$$\vec{A} \rightarrow \vec{A}' = \vec{A} - \vec{\nabla}\chi, \quad A_0 \rightarrow A_0' = A_0 + \frac{\partial \chi}{\partial t} \tag{1.2}$$

where $\chi\left(\vec{x}, t\right)$ is an arbitrary real valued function. Using relativistic notation this can also be written as,

$$A_\nu \rightarrow A_\nu' = A_\nu + \partial_\nu \chi \tag{1.3}$$

For quantum field theory to be gauge invariant a change in the gauge cannot produce a change in any physical observable such as the current and charge expectation values. However it is well known that when certain quantities are calculated using standard perturbation theory the results are not gauge invariant. The non-gauge invariant terms that appear in the results have to be removed to make the answer physically correct.

For example, the first order change in the vacuum current, due to an applied electromagnetic field, can be shown to be given by,

$$J_{vac}^\mu\left(x\right) = \int \pi^{\mu\nu}\left(x - x'\right) A_\nu\left(x'\right) d^4 x' \tag{1.4}$$

where $\pi^{\mu\nu}$ is called the polarization tensor and where, in the above expression, summation over repeated indices is assumed. The above equation is normally written in terms of the Fourier transformed quantities as,

$$J_{vac}^{\mu}(k) = \pi^{\mu\nu}(k) A_{\nu}(k) \tag{1.5}$$

where k is the 4-momentum of the Electromagnetic field. In this case a gauge transformation takes the following form,

$$A_{\nu}(k) \rightarrow A_{\nu}'(k) = A_{\nu}(k) + ik_{\nu}\chi(k) \tag{1.6}$$

The change in the vacuum current, $\delta_g J_{vac}^{\mu}(k)$, due to a gauge transformation can be obtained by using (1.6) in (1.5) to yield,

$$\delta_g J_{vac}^{\mu}(k) = ik_{\nu}\pi^{\mu\nu}(k)\chi(k) \tag{1.7}$$

Now if quantum theory is gauge invariant then an observable quantity, such as the vacuum current, must not be affected by a gauge transformation. Therefore $\delta_g J_{vac}^{\mu}(k)$ must be zero. For this to be true we must have that,

$$k_{\nu}\pi^{\mu\nu}(k) = 0 \tag{1.8}$$

However, when the polarization tensor is calculated it always the case that the above relationship does not hold.

Consider, for example, a calculation of the vacuum polarization tensor by Heitler (see page 322 of [3]). Heitler's solution for the Fourier transform of the vacuum polarization tensor is,

$$\pi^{uv}(k) = \pi_G^{uv}(k) + \pi_{NG}^{uv}(k) \tag{1.9}$$

The first term on the right hand side is given by,

$$\pi_G^{uv}\left(k\right)=\left(\frac{2q^2}{3\pi}\right)\left(k^{\mu}k^{\nu}-g^{\mu\nu}k^2\right)\int\limits_{2m}^{\infty}dz\frac{\left(z^2+2m^2\right)\sqrt{\left(z^2-4m^2\right)}}{z^2\left(z^2-k^2\right)} \tag{1.10}$$

where m is the mass of the electron, q is the electric charge, and $\hbar=c=1$. This term is gauge invariant because $k_{\nu}\pi_G^{uv}=0$. The second term on the right of (1.9) is

$$\pi_{NG}^{uv}\left(k\right)=\left(\frac{2q^2}{3\pi}\right)g_{\nu}^{\mu}\left(1-g^{\mu0}\right)\int\limits_{2m}^{\infty}dz\frac{\left(z^2+2m^2\right)\sqrt{\left(z^2-4m^2\right)}}{z^2} \tag{1.11}$$

where there is no summation over the two $\mu$ superscripts that appear on the right. Note that $\pi_{NG}^{uv}$ is not gauge invariant because $k_{\nu}\pi_{NG}^{uv}\neq 0$. Therefore to get a physically valid result it is necessary to "correct" equation (1.9) by dropping $\pi_{NG}^{uv}$ from the solution.

A similar situation exists when other sources in the literature are examined. For example consider the discussion in Section 14.2 of Greiner et al [2]. Greiner writes the solution for the vacuum polarization tensor (see equation 14.43 of [2]) as,

$$\pi^{\mu\nu}\left(k\right)=\left(g^{\mu\nu}k^2-k^{u}k^{\nu}\right)\pi\left(k^2\right)+g^{\mu\nu}\pi_{sp}\left(k^2\right) \tag{1.12}$$

where the quantities $\pi\left(k^2\right)$ and $\pi_{sp}\left(k^2\right)$ are given in [2]. Referring to (1.8) it can be easily shown that the first term on the right is gauge invariant. However the second term is not gauge invariant unless $\pi_{sp}\left(k^2\right)$ equals zero. Greiner shows that this is not the case. Therefore this term must be dropped form the result in order to obtain a physically valid solution.

For another example of this refer to section 6-4 of Nishijima [4]. In this reference is it shown that the vacuum polarization tensor includes a non-gauge invariant term which must be removed. For other examples refer to equation 7.79 of Peskin and Schroeder [5] and Section 5.2 of Greiner and Reinhardt [6]. In all cases a direct calculation of the vacuum polarization tensor using perturbation theory produces a result which includes non-gauge invariant terms. In all cases the non-gauge invariant terms must be removed to obtain the "correct" gauge invariant result.

There are two general approaches to removing these non-gauge invariant terms. The first approach is simply to recognize that the term cannot be physically valid and drop it from the solution. This is the approach taken by Heitler [3], Nishijima [4], and Greiner et al [2]. The other approach is to come up with mathematical techniques which automatically eliminate the offending terms. This is called "regularization". There are two types of regularization. One type is called Pauli-Villars regularization [7] . In this case additional functions are introduced that have the correct behavior so that the non-gauge invariant terms are cancelled. An example of the use Pauli-Villars regularization is given by Greiner and Reinhardt [6]. Another type of regularization is called dimensional regularization. An example of this is given by Peskin and Schroeder [5].

The problem with regularization is that there is no physical explanation for why it is required. It is an ad hoc mathematical device required to remove the unwanted terms. Also regularization does not guarantee a unique answer. For example consider the result from [2] given by equation (1.12). As discussed above the second term on the right must be removed. In reference [2] the authors simply removed the term $g^{\mu\nu}\pi_{sp}\left(k^2\right)$ by hand without resorting to formal regularization schemes. Suppose someone invented a mathematical procedure that subtracted the term $g^{\mu\nu}\pi_{sp}\left(k^2\right)+\left(g^{\mu\nu}k^2-k^uk^v\right)f\left(k^2\right)$ from (1.12) where $f\left(k^2\right)$ is some arbitrary function. The final "corrected" result would be $\left(g^{\mu\nu}k^2 - k^u k^v\right)\left(\pi\left(k^2\right) - f\left(k^2\right)\right)$.  Since this is gauge invariant it would certainly be a physically acceptable solution, however it is not unique because $f\left(k^2\right)$ is arbitrary.  That is, there is no theoretical way to distinguish between a mathematical regularization procedure for which $f\left(k^2\right) = 0$ and one for which $f\left(k^2\right)$ is non-zero.  Both yield physically valid results.

The obvious question that should be asked is why is regularization needed at all?  That is, why do non-gauge invariant terms appear in a theory that is supposed to be gauge invariant?  This question is briefly addressed by Greiner et al (page 398 of [2]) who writes "… this latter [non-gauge invariant] term violates the gauge invariance of the theory.  This is a very sever contradiction to the experimentally confirmed gauge independence of QED.  [This problem indicates] that perturbative QED is not a complete theory.  As one counter example or inconsistency suffices to prove a theory wrong, we should, in principle, spend the rest of this book searching for an improved theory.  However, there is little active work on this today because: (1) there is a common belief that some artifact of the exact mathematics is the source of the problem; (2) this problem may disappear when the properly generalized theory, including in its framework all charged Dirac particles, is achieved."

It is my impression that the above paragraph pretty much expresses the current state of thinking on this problem, i.e. , the problem is probably due to an artifact of the mathematics and will, hopefully, go away when a complete theory is revealed.  It is the purpose of this paper to more fully address the question on why these non-gauge invariant terms appear.  It will be shown that this is not an artifact of the mathematics but is a result of the underlying structure of Dirac field theory and, in particular, the way the vacuum state is defined.

## II. Dirac field theory

In this section the basic elements of Dirac field theory in the Schrödinger representation will be introduced. Natural unit will be used so that $\hbar = c = 1$. In the Schrödinger representation of Dirac field theory the time evolution of the state vector $\left| \Omega(t) \right\rangle$ and its dual $\left\langle \Omega(t) \right|$ are given by the Schrödinger equation,

$$\frac{\partial \left| \Omega(t) \right\rangle}{\partial t} = -i\hat{H}(t) \left| \Omega(t) \right\rangle, \quad \frac{\partial \left\langle \Omega(t) \right|}{\partial t} = i \left\langle \Omega(t) \right| \hat{H}(t) \tag{2.1}$$

The Hamiltonian operator $\hat{H}(t)$ is given by (see Chapt. 9 of [2]),

$$\hat{H}(t) = \frac{1}{2} \int \left[ \hat{\psi}^\dagger(\vec{x}), H(t)\hat{\psi}(\vec{x}) \right] d\vec{x} \tag{2.2}$$

where if $\hat{O}$ is some operator then,

$$\left[ \hat{\psi}^\dagger, \hat{O}\hat{\psi} \right] \equiv \hat{\psi}_\alpha^\dagger \left( \hat{O}_{\alpha\beta}\hat{\psi}_\beta \right) - \left( \hat{O}_{\alpha\beta}\hat{\psi}_\beta \right) \hat{\psi}_\alpha^\dagger \tag{2.3}$$

and where,

$$H(t) = H_0 - q\vec{\alpha} \cdot \vec{A}(\vec{x},t) + qA_0(\vec{x},t) \tag{2.4}$$

and

$$H_0 = -i\vec{\alpha} \cdot \vec{\nabla} + \beta m \tag{2.5}$$

In the above expression q is the electric charge, m is the mass of the electron, $\vec{\alpha}$ and $\beta$ are the usual 4x4 matrices (see [2]), and $\hat{\psi}(\vec{x})$ is the field operator. Note that the field operator is time independent. This is consistent with the Schrödinger representation where the field operators are time independent and the time dependence of the system is carried by the state vector. It is convenient to rewrite (2.2) as,

$$\hat{H}(t) = \hat{H}_0 - \int \hat{\vec{J}}(\vec{x}) \cdot \vec{A}(\vec{x},t) d\vec{x} + \int \hat{\rho}(\vec{x}) A_0(\vec{x},t) d\vec{x} \qquad (2.6)$$

where,

$$\hat{H}_0 = \frac{1}{2} \int \left[ \hat{\psi}^\dagger(\vec{x}), H_0 \hat{\psi} \right] (\vec{x}) d\vec{x} \qquad (2.7)$$

and

$$\hat{\vec{J}}(\vec{x}) = \frac{q}{2} \left[ \hat{\psi}^\dagger(\vec{x}), \vec{\alpha} \hat{\psi}(\vec{x}) \right]; \quad \hat{\rho}(\vec{x}) = \frac{q}{2} \left[ \hat{\psi}^\dagger(\vec{x}), \hat{\psi}(\vec{x}) \right] \qquad (2.8)$$

In the above expressions $\hat{\vec{J}}(\vec{x})$ is the current operator, $\hat{\rho}(\vec{x})$ is the charge operator, and $\hat{H}_0$ is the free field Hamiltonian which is the Hamiltonian when the electric potential is zero. The field operator is given by,

$$\hat{\psi}(\vec{x}) = \sum_n \hat{a}_n \phi_n(\vec{x}); \quad \hat{\psi}^\dagger(\vec{x},t) = \sum_n \hat{a}_n^\dagger \phi_n^\dagger(\vec{x}) \qquad (2.9)$$

where the $\hat{a}_n$ ($\hat{a}_n^\dagger$) are the destruction(creation) operators for a particle in the state $\phi_n(\vec{x})$. They satisfy the anticommutator relation,

$$\hat{a}_m \hat{a}_n^\dagger + \hat{a}_n^\dagger \hat{a}_m = \delta_{mn}; \text{ all other anticommutators=0} \qquad (2.10)$$

The $\phi_n(\vec{x})$ are basis state solutions of the free field Dirac equation with energy eigenvalue $\lambda_n E_n$ and can be expressed by

$$H_0 \phi_n(\vec{x}) = \lambda_n E_n \phi_n(\vec{x}) \qquad (2.11)$$

and where,

$$E_n = +\sqrt{\vec{p}_n^2 + m^2}, \quad \lambda_n = \begin{cases} +1 \text{ for a positive energy state} \\ -1 \text{ for a negative energy state} \end{cases} \qquad (2.12)$$

where $\vec{p}_n$ is the momentum of the state n. The $\phi_n(\vec{x})$ can be expressed by,

$$\phi_n(\vec{x}) = u_n e^{i\vec{p}\cdot\vec{x}} \tag{2.13}$$

where $u_n$ is a constant 4-spinor (see [2]). The $\phi_n(\vec{x})$ form a complete orthonormal basis in Hilbert space and satisfy

$$\int \phi_n^\dagger(\vec{x})\phi_m(\vec{x})d\vec{x} = \delta_{mn} \tag{2.14}$$

and

$$\sum_n \left(\phi_n^\dagger(\vec{x})\right)_{(a)}\left(\phi_n(\vec{y})\right)_{(b)} = \delta_{ab}\delta^3(\vec{x}-\vec{y}) \tag{2.15}$$

where "a" and "b" are spinor indices (see page 107 of Heitler [3]).

Following Greiner [2] define the state vector $|0, \text{bare}\rangle$ which is the state vector that is empty of all particles, i.e.,

$$\hat{a}_n|0,\text{bare}\rangle = 0 \text{ for all n} \tag{2.16}$$

Order the index 'n' associated with the basis wave function $\phi_n^{(0)}(\vec{x})$ so that $n > 0$ implies a positive energy state and $n < 0$ implies a negative energy state. Further order the states so that the magnitude of 'n' increases with the magnitude of the energy. The vacuum state is, then, defined by,

$$|0\rangle = \left(\hat{a}_{-1}^\dagger \hat{a}_{-2}^\dagger \hat{a}_{-3}^\dagger \ldots\right)|0,\text{bare}\rangle = \prod_{n=1}^\infty \hat{a}_{-n}^\dagger|0,\text{bare}\rangle \tag{2.17}$$

At this point we can think of the vacuum as consisting of an infinite number of negative energy electrons. This definition can be simplified by replacing the operator $\hat{a}_{-j}$ ($j > 0$), that destroys a negative energy electron, by the operator $\hat{d}_j^\dagger$, that creates a positive energy positron. Thus we make the following change in notation,

$$\hat{b}_j = \hat{a}_j \rightarrow \hat{b}_j^\dagger = \hat{a}_j^\dagger \text{ and } \hat{d}_j = \hat{a}_{-j}^\dagger \rightarrow \hat{d}_j^\dagger = \hat{a}_{-j} \text{ where } j > 0 \tag{2.18}$$

In the above expressions $\hat{d}_j$ and $\hat{d}_j^\dagger$ are positron destruction and creation operators, respectively. The $\hat{b}_j$ and $\hat{b}_j^\dagger$ are the electron destruction and creation operators, respectively, that act on electrons in positive energy states. Use this notation in (2.9) to obtain,

$$\hat{\psi}(\bar{x}) = \sum_{j=1}^{\infty} \left( \hat{b}_j \phi_j(\bar{x}) + \hat{d}_j^\dagger \phi_{-j}(\bar{x}) \right) \tag{2.19}$$

and,

$$\hat{\psi}^\dagger(\bar{x}) = \sum_{j=1}^{\infty} \left( \hat{b}_j^\dagger \phi_j^\dagger(\bar{x}) + \hat{d}_j \phi_{-j}^\dagger(\bar{x}) \right) \tag{2.20}$$

These operators satisfy,

$$\left\{ \hat{d}_j, \hat{d}_k^\dagger \right\} = \delta_{jk} \text{ ; } \left\{ \hat{b}_j, \hat{b}_k^\dagger \right\} = \delta_{jk} \text{ ; all other anti-commutators are zero} \tag{2.21}$$

The vacuum state $|0\rangle$ is now defined by,

$$\hat{d}_j|0\rangle = \hat{b}_j|0\rangle = 0 \text{ and } \langle 0|\hat{d}_j^\dagger = \langle 0|\hat{b}_j^\dagger = 0 \tag{2.22}$$

Using the above results the free field Hamiltonian operator becomes,

$$\hat{H}_0 = \frac{1}{2} \sum_{n=1}^{\infty} \left( \lambda_n E_n \left( \hat{b}_n^\dagger \hat{b}_n - \hat{b}_n \hat{b}_n^\dagger \right) + \lambda_{-n} E_{-n} \left( \hat{d}_n \hat{d}_n^\dagger - \hat{d}_n^\dagger \hat{d}_n \right) \right) \tag{2.23}$$

Use (2.21) in the above to yield,

$$\hat{H}_0 = \sum_{n=1}^{\infty} \left( E_n \hat{b}_n^\dagger \hat{b}_n + E_n \hat{d}_n^\dagger \hat{d}_n \right) - E_{vac} \tag{2.24}$$

where we have used $E_n = E_{-n}$, $\lambda_n = 1$, and $\lambda_{-n} = -1$ for $n > 0$ and,

$$E_{vac} = \sum_{n=1}^{\infty} E_n \qquad (2.25)$$

At this point redefine $\hat{H}_0$ by adding the constant $E_{vac}$ to obtain,

$$\hat{H}_0 = \sum_{n=1}^{\infty} \left( E_n \hat{b}_n^{\dagger} \hat{b}_n + E_n \hat{d}_n^{\dagger} \hat{d}_n \right) \qquad (2.26)$$

This last step does not affect any results and simply corresponds to a shift in energy, making the energy of the vacuum state equal to zero. This will simplify some of the mathematical analysis but does not change any of the results of the following discussion.

We can now think of the vacuum state $|0\rangle$ as the state which contains no electrons, no positrons, and has zero energy. New eigenstates states $|k\rangle$ are created by acting on $|0\rangle$ with electron and positron creation operators, e.g,

$$|k\rangle = \hat{b}_{j_1}^{\dagger} \hat{b}_{j_2}^{\dagger} ... \hat{b}_{j_s}^{\dagger} \hat{d}_{v_1}^{\dagger} \hat{d}_{v_2}^{\dagger} ... \hat{d}_{v_r}^{\dagger} |0\rangle \qquad (2.27)$$

From this definition and (2.26) we have,

$$\hat{H}_0 |k\rangle = \xi_{|k\rangle} |k\rangle \qquad (2.28)$$

where $\xi_{|k\rangle}$ is the energy eigenvalue of the eigenstate $|k\rangle$ and is given by,

$$\xi_{|k\rangle} = \left( E_{j_1} + E_{j_2} + ... + E_{j_s} \right) + \left( E_{v_1} + E_{v_2} + ... + E_{v_r} \right) \qquad (2.29)$$

Since all the quantities $E_{j_i}$ and $E_{v_i}$ in the above equation are greater than zero we have that,

$$\xi_{|k\rangle} > \xi_{|0\rangle} = 0 \text{ for all } |k\rangle \neq |0\rangle \qquad (2.30)$$

The eigenstates $|k\rangle$ form an orthonormal set in fock space and satisfy [8],

$$\langle k | q \rangle = \delta_{kq} \text{ and } \sum_{|k\rangle} |k\rangle\langle k| = 1 \qquad (2.31)$$

where the summation is over all eigenstates $|k\rangle$. Any arbitrary state vector $|\Omega\rangle$ can be expressed as an expansion in terms of the eigenstates $|k\rangle$ as,

$$|\Omega\rangle = \sum_{|k\rangle} c_{|k\rangle} |k\rangle \qquad (2.32)$$

The free field energy of a given normalized state $|\Omega\rangle$ is defined by,

$$\xi_f\left(|\Omega\rangle\right) = \langle\Omega|\hat{H}_0|\Omega\rangle \qquad (2.33)$$

From the above discussion we have that,

$$\xi_f\left(|\Omega\rangle\right) = \langle\Omega|\hat{H}_0|\Omega\rangle = \sum_{|k\rangle} \left|c_{|k\rangle}\right|^2 \xi_{|k\rangle} \qquad (2.34)$$

Since the $\xi_{|k\rangle}$ are all positive we have that,

$$\xi_f\left(|\Omega\rangle\right) \geq \xi_f\left(|0\rangle\right) = 0 \text{ for all } |\Omega\rangle \qquad (2.35)$$

Therefore the vacuum state $|0\rangle$ represents a lower bound to the free field energy.

### III. The problem of gauge invariance

The discussion in the last section has been a review of various elements of Dirac field theory. These elements, in one form or another, are found in many basic works on quantum mechanics. In this section we will consider the following problem – are the above elements of Dirac field theory compatible with principle of gauge invariance? The basic elements of the following discussion are also presented in [9].

Consider a normalized state vector $|\Omega(t)\rangle$. Assume that at some initial time, say $t = t_i$, $|\Omega(t_i)\rangle$ satisfies (2.35), i.e., $\xi_f\left(|\Omega(t_i)\rangle\right) \geq 0$. Now let $|\Omega(t)\rangle$ evolve according

to the Schrödinger equation (2.1). If the electric potential is non-zero then, in general, the free field energy $\xi_f\left(\left|\Omega(t)\right\rangle\right)$ will change in time. It will be shown that if we assume that field theory is gauge invariant then it is possible to specify an electric potential so that (2.35) is not true at some final time $t_f > t_i$. Therefore if (2.35) is true then the principle of gauge invariance is not valid for Dirac field theory.

Start by taking the time derivative of $\xi_f\left(\left|\Omega(t)\right\rangle\right)$ and use (2.1) and (2.33) to obtain,

$$\frac{\partial \xi_f\left(\left|\Omega(t)\right\rangle\right)}{\partial t} = i\left\langle\Omega(t)\right|\left[\hat{H},\hat{H}_0\right]\left|\Omega(t)\right\rangle \tag{3.1}$$

Next use (2.6) and substitute for $\hat{H}_0$ in the above to obtain,

$$\frac{\partial \xi_f\left(\left|\Omega(t)\right\rangle\right)}{\partial t} = i\left\langle\Omega(t)\right|\left[\hat{H},\left(\hat{H}+\int\hat{\vec{J}}(\vec{x})\cdot\vec{A}(\vec{x},t)d\vec{x}-\int\hat{\rho}(\vec{x})A_0(\vec{x},t)d\vec{x}\right)\right]\left|\Omega(t)\right\rangle \tag{3.2}$$

Rearrange terms to yield,

$$\frac{\partial \xi_f\left(\left|\Omega(t)\right\rangle\right)}{\partial t} = i\left(\begin{array}{c}\int\left\langle\Omega(t)\right|\left[\hat{H},\hat{\vec{J}}(\vec{x})\right]\left|\Omega(t)\right\rangle\cdot\vec{A}(\vec{x},t)d\vec{x}\\-\int\left\langle\Omega(t)\right|\left[\hat{H},\hat{\rho}(\vec{x})\right]\left|\Omega(t)\right\rangle A_0(\vec{x},t)d\vec{x}\end{array}\right) \tag{3.3}$$

The current and charge expectation values are defined by,

$$\vec{J}_e(\vec{x},t) = \left\langle\Omega(t)\right|\hat{\vec{J}}(\vec{x})\left|\Omega(t)\right\rangle \text{ and } \rho_e(\vec{x},t) = \left\langle\Omega(t)\right|\hat{\rho}(\vec{x})\left|\Omega(t)\right\rangle \tag{3.4}$$

Use this along with (2.1) to obtain,

$$\frac{\partial \vec{J}_e(\vec{x},t)}{\partial t} = i\left\langle\Omega(t)\right|\left[\hat{H},\hat{\vec{J}}(\vec{x})\right]\left|\Omega(t)\right\rangle \text{ and } \frac{\partial \rho_e(\vec{x},t)}{\partial t} = i\left\langle\Omega(t)\right|\left[\hat{H},\hat{\rho}(\vec{x})\right]\left|\Omega(t)\right\rangle \tag{3.5}$$

Use this result in (3.3) to yield,

$$\frac{\partial \xi_f \left( \left| \Omega(t) \right\rangle \right)}{\partial t} = \left( \int \frac{\partial \vec{J}_e \left( \vec{x}, t \right)}{\partial t} \cdot \vec{A} \left( \vec{x}, t \right) d\vec{x} - \int \frac{\partial \rho_e \left( \vec{x}, t \right)}{\partial t} A_0 \left( \vec{x}, t \right) d\vec{x} \right) \qquad (3.6)$$

Now let the electric potential be given by,

$$\left( A_0 \left( \vec{x}, t \right), \vec{A} \left( \vec{x}, t \right) \right) = \left( \frac{\partial \chi \left( \vec{x}, t \right)}{\partial t}, -\vec{\nabla} \chi \left( \vec{x}, t \right) \right) \qquad (3.7)$$

where $\chi \left( \vec{x}, t \right)$ is an arbitrary real valued function with the initial condition, at $t = t_i$,

$$\frac{\partial \chi \left( \vec{x}, t_i \right)}{\partial t_i} = 0 \text{ and } \chi \left( \vec{x}, t_i \right) = 0 \qquad (3.8)$$

Use this in (3.6) to obtain,

$$\frac{\partial \xi_f \left( \left| \Omega(t) \right\rangle \right)}{\partial t} = \left( -\int \frac{\partial \vec{J}_e \left( \vec{x}, t \right)}{\partial t} \cdot \vec{\nabla} \chi \left( \vec{x}, t \right) d\vec{x} - \int \frac{\partial \rho_e \left( \vec{x}, t \right)}{\partial t} \frac{\partial \chi \left( \vec{x}, t \right)}{\partial t} d\vec{x} \right) \qquad (3.9)$$

Integrate the first integral by parts and assume reasonable boundary conditions to obtain,

$$\frac{\partial \xi_f \left( \left| \Omega(t) \right\rangle \right)}{\partial t} = \left( \int \frac{\partial \vec{\nabla} \cdot \vec{J}_e \left( \vec{x}, t \right)}{\partial t} \chi \left( \vec{x}, t \right) d\vec{x} - \int \frac{\partial \rho_e \left( \vec{x}, t \right)}{\partial t} \frac{\partial \chi \left( \vec{x}, t \right)}{\partial t} d\vec{x} \right) \qquad (3.10)$$

This can be written as,

$$\frac{\partial \xi_f \left( \left| \Omega(t) \right\rangle \right)}{\partial t} = \int \chi \left( \vec{x}, t \right) \frac{\partial}{\partial t} \left( \frac{\partial \rho_e \left( \vec{x}, t \right)}{\partial t} + \vec{\nabla} \cdot \vec{J}_e \left( \vec{x}, t \right) \right) d\vec{x} - \frac{\partial}{\partial t} \int \frac{\partial \rho_e \left( \vec{x}, t \right)}{\partial t} \chi \left( \vec{x}, t \right) d\vec{x} \qquad (3.11)$$

Integrate the above with respect to time from the initial time $t = t_i$ to some final time

$t_f > t_i$ and use (3.8) to obtain,

$$\xi_f \left( \left| \Omega(t_f) \right\rangle \right) - \xi_f \left( \left| \Omega(t_i) \right\rangle \right) = \int_{t_i}^{t_f} dt \left( \int \chi \left( \vec{x}, t \right) \frac{\partial L \left( \vec{x}, t \right)}{\partial t} d\vec{x} \right) - \int \left( \frac{\partial \rho_e \left( \vec{x}, t_f \right)}{\partial t_f} \right) \chi \left( \vec{x}, t_f \right) d\vec{x}$$

$$(3.12)$$

where,

$$L(\vec{x}, t) \equiv \frac{\partial \rho_e(\vec{x}, t)}{\partial t} + \vec{\nabla} \cdot \vec{J}_e(\vec{x}, t) \qquad (3.13)$$

Next invoke the principle of gauge invariance. When the electric potential of (3.7) is substituted into (1.1) it can be seen that the electromagnetic field is zero for all functions $\chi(\vec{x}, t)$. Now the current expectation value, $\vec{J}_e(\vec{x}, t)$, and the charge expectation value, $\rho_e(\vec{x}, t)$, are physical observables. Therefore, from the principle of gauge invariance, these quantities are independent of $\chi(\vec{x}, t)$. Therefore $L(\vec{x}, t)$ is independent of $\chi(\vec{x}, t)$. This means that $\chi(\vec{x}, t)$ can be varied in an arbitrary manner without affecting $\vec{J}_e(\vec{x}, t)$, $\rho_e(\vec{x}, t)$, or $L(\vec{x}, t)$. Note that $L(\vec{x}, t) = 0$ is the continuity equation. Since local charge conservation is an experimental fact we could simplify the above equation by using $L(\vec{x}, t) = 0$. However the proof is not dependent on the validity of the continuity equation therefore the following two possibilities will be considered. In first case we will let $L(\vec{x}, t) = 0$. Use this in (3.12) to obtain,

$$\xi_f\left(\left|\Omega(t_f)\right\rangle\right) = \xi_f\left(\left|\Omega(t_i)\right\rangle\right) - \int \chi(\vec{x}, t_f) \frac{\partial \rho_e(\vec{x}, t_f)}{\partial t_f} d\vec{x} \qquad (3.14)$$

Now since $\chi$ is arbitrary it is always possible to find a $\chi$ which makes $\xi_f\left(\left|\Omega(t_f)\right\rangle\right)$ a negative number with an arbitrary large magnitude. For example let $\chi(\vec{x}, t_f) = f \frac{\partial \rho_e(\vec{x}, t_f)}{\partial t_f}$ where f is a constant. Use this in (3.14) to obtain,

$$\xi_f\left(\left|\Omega(t_f)\right\rangle\right) = \xi_f\left(\left|\Omega(t_i)\right\rangle\right) - f \int \left(\frac{\partial \rho_e(\vec{x}, t_f)}{\partial t_f}\right)^2 d\vec{x} \qquad (3.15)$$

Assume that we have selected a quantum state so that $\dfrac{\partial \rho_e\left(\vec{x}, t_f\right)}{\partial t_f} \neq 0$. In this case the integral is positive. Therefore as $f \to \infty$ we have that $\xi_f\left(\left|\Omega\left(t_f\right)\right\rangle\right) \to -\infty$. Now how do we know that the condition $\dfrac{\partial \rho_e\left(\vec{x}, t_f\right)}{\partial t_f} \neq 0$ can be satisfied. If quantum mechanics is correct model of the real world then there must be quantum states which meet this condition because there are many situations in the real world where $\dfrac{\partial \rho_e\left(\vec{x}, t\right)}{\partial t} \neq 0$.

Next consider the case where $\partial L\left(\vec{x}, t\right) / \partial t \neq 0$. In this case we can set,

$$\chi\left(\vec{x}, t\right) = \left\langle \begin{array}{l} -f \dfrac{\partial L\left(\vec{x}, t\right)}{\partial t} \text{ for } t_i < t < t_f \\[2mm] 0 \text{ for } t \leq t_i \text{ and } t \geq t_f \end{array} \right. \tag{3.16}$$

Use this in (3.12) to obtain,

$$\xi_f\left(\left|\Omega\left(t_f\right)\right\rangle\right) = \xi_f\left(\left|\Omega\left(t_i\right)\right\rangle\right) - f \int_{t_i}^{t_f} dt \int \left(\dfrac{\partial L\left(\vec{x}, t\right)}{\partial t}\right)^2 d\vec{x} \tag{3.17}$$

Once again the integral is always positive so that as $f \to \infty$ then $\xi_f\left(\left|\Omega\left(t_f\right)\right\rangle\right) \to -\infty$.

What we have shown is that if Dirac field theory is gauge invariant there must exist quantum states $\left|\Omega\left(t_f\right)\right\rangle$ whose free field energy approaches negative infinity. That is there is no lower bound to the free field energy. This result is in direct contradiction to the relationship given by equation (2.35).

To summarize the results of this section we have used the Schrödinger equation (2.1) and the definition of the Hamiltonian operator (2.6) to obtain an expression for the first derivative of the free field energy per equation (3.2). We evaluate this expression in the presence of the electric potential given by (3.7) to obtain (3.11). This is then integrated with respect to time to obtain (3.12). At this point the principle of gauge invariance is invoked. This allows us to vary $\chi$ in an arbitrary matter without effecting the observables $\rho_e\left(\vec{x}, t\right)$ and $\vec{J}_e\left(\vec{x}, t\right)$. Therefore we can find a $\chi$ for which

$$\xi_f\left(\left|\Omega\left(t_f\right)\right\rangle\right) \rightarrow -\infty$$ in direct contradiction to (2.35), where equation (2.35) is derived from the standard definition of the vacuum state $|0\rangle$.

The result of this analysis is the standard elements of Dirac field theory, which are presented in Section II, are not consistent with the principle of gauge invariance. Therefore standard Dirac field theory is *not* gauge invariant. This, then, is why calculations in field theory do not directly yield gauge invariant result. The underlying theory is not gauge invariant due to the fact that there is a lower bound to the free field energy as expressed by (2.35).

## IV. The Vacuum Current

In this section we will show why a calculation of the vacuum current produces a non-gauge invariant result. The first order change in the vacuum current due to an applied electric potential is given by the following expression which is derived in Appendix A (see also [9] and Eq. 8.3 of [10] ).

$$\vec{J}_{vac}\left(\vec{x}, t\right) = i\left\langle 0\left|\left[\hat{\vec{J}}_I\left(\vec{x}, t\right), \int d\vec{y}\int_{-\infty}^{t} dt'\left(-\hat{\vec{J}}_I\left(\vec{y}, t'\right) \cdot \vec{A}\left(\vec{y}, t'\right) + \hat{\rho}_I\left(\vec{y}, t'\right)A_o\left(\vec{y}, t'\right)\right)\right]\right|0\right\rangle \quad (4.1)$$

where the operators $\hat{\vec{J}}_I\left(\vec{x}, t\right)$ and $\hat{\rho}_I\left(\vec{x}, t\right)$ are the current and charge operators, respectively, in the interaction representation. They are defined by,

$$\hat{\vec{J}}_I\left(\vec{x}, t\right) = e^{i\hat{H}_0 t}\hat{\vec{J}}\left(\vec{x}\right)e^{-i\hat{H}_0 t} \text{ and } \hat{\rho}_I\left(\vec{x}, t\right) = e^{i\hat{H}_0 t}\hat{\rho}\left(\vec{x}\right)e^{-i\hat{H}_0 t} \quad (4.2)$$

They can be shown to satisfy the continuity equation (see Eq. 3.11 of [10] and Appendix B ),

$$\frac{\partial \hat{\rho}_I\left(\vec{x},t\right)}{\partial t} = -\vec{\nabla} \cdot \hat{\vec{J}}_I\left(\vec{x},t\right) \tag{4.3}$$

The change in the vacuum current $\delta_g \vec{J}_{vac}\left(\vec{x},t\right)$ due to a gauge transformation is obtained by using (1.2) in (4.1) to yield

$$\delta_g \vec{J}_{vac}\left(\vec{x},t\right) = i\left\langle 0\left|\left[\hat{\vec{J}}_I\left(\vec{x},t\right), \int d\bar{y}\int_{-\infty}^{t} dt'\left(\hat{\vec{J}}_I\left(\vec{y},t'\right)\cdot\vec{\nabla}\chi\left(\vec{y},t'\right) + \hat{\rho}_I\left(\vec{y},t'\right)\frac{\partial\chi\left(\vec{y},t'\right)}{\partial t'}\right)\right]\right|0\right\rangle \tag{4.4}$$

If quantum field theory is gauge invariant then a gauge transformation should produce no change in any observable quantity. Therefore $\delta_g \vec{J}_{vac}\left(\vec{x},t\right)$ should be zero. To see if this is the case we will solve the above equation as follows. First consider the following relationship,

$$\int_{-\infty}^{t} dt'\hat{\rho}_I\left(\vec{y},t'\right)\frac{\partial\chi\left(\vec{y},t'\right)}{\partial t'} = \Big|_{-\infty}^{t}\hat{\rho}_I\left(\vec{y},t'\right)\chi\left(\vec{y},t'\right) - \int_{-\infty}^{t} dt'\chi\left(\vec{y},t'\right)\frac{\partial\hat{\rho}_I\left(\vec{y},t'\right)}{\partial t'} \tag{4.5}$$

Assume that $\chi\left(\vec{y},t\right) = 0$ at $t \to -\infty$. Use this and (4.3) in the above expression to obtain

$$\int_{-\infty}^{t} dt'\hat{\rho}_I\left(\vec{y},t'\right)\frac{\partial\chi\left(\vec{y},t'\right)}{\partial t'} = \hat{\rho}_I\left(\vec{y},t\right)\chi\left(\vec{y},t\right) + \int_{-\infty}^{t} dt'\chi\left(\vec{y},t'\right)\vec{\nabla}\cdot\vec{J}_I\left(\vec{y},t'\right) \tag{4.6}$$

Substitute this into (4.4) to obtain

$$\delta_g \vec{J}_{vac}\left(\vec{x},t\right) = i\left\langle 0\left|\left[\hat{\vec{J}}_I\left(\vec{x},t\right), \int d\bar{y}\int_{-\infty}^{t} dt'\left(\hat{\vec{J}}_I\left(\vec{y},t'\right)\cdot\vec{\nabla}\chi\left(\vec{y},t'\right) + \chi\left(\vec{y},t'\right)\vec{\nabla}\cdot\hat{\vec{J}}_I\left(\vec{y},t'\right)\right)\right]\right|0\right\rangle$$
$$+ i\left\langle 0\left|\left[\hat{\vec{J}}_I\left(\vec{x},t\right), \int \hat{\rho}_I\left(\vec{y},t\right)\chi\left(\vec{y},t\right)d\bar{y}\right]\right|0\right\rangle \tag{4.7}$$

Rearrange terms to obtain

$$\delta_g \vec{J}_{vac}(\vec{x}, t) = i \left\langle 0 \left| \left[ \hat{\vec{J}}_I(\vec{x}, t), \int_{-\infty}^{t} dt' \int d\vec{y} \vec{\nabla} \cdot \left( \hat{\vec{J}}_I(\vec{y}, t') \chi(\vec{y}, t') \right) \right] \right| 0 \right\rangle$$
$$+ i \left\langle 0 \left| \left[ \hat{\vec{J}}_I(\vec{x}, t), \int \hat{\rho}_I(\vec{y}, t) \chi(\vec{y}, t) d\vec{y} \right] \right| 0 \right\rangle \quad (4.8)$$

Assume reasonable boundary conditions at $|\vec{y}| \to \infty$ so that

$$\int d\vec{y} \vec{\nabla} \cdot \left( \hat{\vec{J}}_I(\vec{y}, t') \chi(\vec{y}, t') \right) = 0 \quad (4.9)$$

Use this in (4.8) to obtain

$$\delta_g \vec{J}_{vac}(\vec{x}, t) = i \left\langle 0 \left| \left[ \hat{\vec{J}}_I(\vec{x}, t), \int \hat{\rho}_I(\vec{y}, t) \chi(\vec{y}, t) d\vec{y} \right] \right| 0 \right\rangle$$
$$= i \int \left\langle 0 \left| \left[ \hat{\vec{J}}_I(\vec{x}, t), \hat{\rho}_I(\vec{y}, t) \right] \right| 0 \right\rangle \chi(\vec{y}, t) d\vec{y} \quad (4.10)$$

Use (4.2) and the fact that $H_0 |0\rangle = \langle 0| H_0 = 0$ to show that,

$$\left\langle 0 \left| \left[ \hat{\vec{J}}_I(\vec{x}, t), \hat{\rho}_I(\vec{y}, t) \right] \right| 0 \right\rangle = \left\langle 0 \left| \left[ \hat{\vec{J}}(\vec{x}), \hat{\rho}(\vec{y}) \right] \right| 0 \right\rangle \quad (4.11)$$

where $\left[ \hat{\vec{J}}(\vec{x}), \hat{\rho}(\vec{y}) \right]$ is commonly called the Schwinger term. Use this in (4.10) to obtain,

$$\delta_g \vec{J}_{vac}(\vec{x}, t) = i \int \left\langle 0 \left| \left[ \hat{\vec{J}}(\vec{x}), \hat{\rho}(\vec{y}) \right] \right| 0 \right\rangle \chi(\vec{y}, t) d\vec{y} \quad (4.12)$$

Therefore for $\delta_g \vec{J}_{vac}(\vec{x}, t)$ to be zero, for arbitrary $\chi(\vec{y}, t)$, the quantity

$\left\langle 0 \left| \left[ \hat{\vec{J}}(\vec{x}), \hat{\rho}(\vec{y}) \right] \right| 0 \right\rangle$ must be zero. However it has been shown by Schwinger [11] that

this term cannot be zero. To prove this refer to Appendix B where we prove the relationship,

$$i \left[ \hat{H}_0, \hat{\rho}(\vec{x}) \right] = -\vec{\nabla} \cdot \hat{\vec{J}}(\vec{x}) \quad (4.13)$$

Next evalate,

$$\vec{\nabla}_{\vec{x}} \cdot \langle 0 | \left[ \hat{\vec{J}}(\vec{x}), \hat{\rho}(\vec{y}) \right] | 0 \rangle = \langle 0 | \left[ \vec{\nabla} \cdot \hat{\vec{J}}(\vec{x}), \hat{\rho}(\vec{y}) \right] | 0 \rangle \tag{4.14}$$

Use (4.13) in the above to obtain,

$$i\vec{\nabla}_{\vec{x}} \cdot \langle 0 | \left[ \hat{\vec{J}}(\vec{x}), \hat{\rho}(\vec{y}) \right] | 0 \rangle = \langle 0 | \left[ \left[ \hat{H}_0, \hat{\rho}(\vec{x}) \right], \hat{\rho}(\vec{y}) \right] | 0 \rangle$$

Next expand the commutator to yield,

$$i\vec{\nabla}_{\vec{x}} \cdot \langle 0 | \left[ \hat{\vec{J}}(\vec{x}), \hat{\rho}(\vec{y}) \right] | 0 \rangle = \langle 0 | \begin{pmatrix} \hat{H}_0 \hat{\rho}(\vec{x}) \hat{\rho}(\vec{y}) - \hat{\rho}(\vec{x}) \hat{H}_0 \hat{\rho}(\vec{y}) \\ -\hat{\rho}(\vec{y}) \hat{H}_0 \hat{\rho}(\vec{x}) + \hat{\rho}(\vec{y}) \hat{\rho}(\vec{x}) \hat{H}_0 \end{pmatrix} | 0 \rangle \tag{4.15}$$

Use $\hat{H}_0 | 0 \rangle = 0$ and $\langle 0 | \hat{H}_0 = 0$ to obtain,

$$-i\vec{\nabla}_{\vec{x}} \cdot \langle 0 | \left[ \hat{\vec{J}}(\vec{x}), \hat{\rho}(\vec{y}) \right] | 0 \rangle = \langle 0 | \hat{\rho}(\vec{x}) \hat{H}_0 \hat{\rho}(\vec{y}) | 0 \rangle + \langle 0 | \hat{\rho}(\vec{y}) \hat{H}_0 \hat{\rho}(\vec{x}) | 0 \rangle \tag{4.16}$$

Next set $\vec{y} = \vec{x}$ to obtain,

$$-i\vec{\nabla}_{\vec{x}} \cdot \langle 0 | \left[ \hat{\vec{J}}(\vec{x}), \hat{\rho}(\vec{y}) \right] | 0 \rangle \Big|_{\vec{y}=\vec{x}} = 2 \langle 0 | \hat{\rho}(\vec{x}) \hat{H}_0 \hat{\rho}(\vec{x}) | 0 \rangle \tag{4.17}$$

Use (2.31) in the above to obtain,

$$-i\vec{\nabla}_{\vec{x}} \cdot \langle 0 | \left[ \hat{\vec{J}}(\vec{x}), \hat{\rho}(\vec{y}) \right] | 0 \rangle \Big|_{\vec{y}=\vec{x}} = 2 \sum_{|n\rangle,|m\rangle} \langle 0 | \hat{\rho}(\vec{x}) | n \rangle \langle n | \hat{H}_0 | m \rangle \langle m | \hat{\rho}(\vec{x}) | 0 \rangle \tag{4.18}$$

Next use (2.31) and (2.28) to obtain,

$$-i\vec{\nabla}_{\vec{x}} \cdot \langle 0 | \left[ \hat{\vec{J}}(\vec{x}), \hat{\rho}(\vec{y}) \right] | 0 \rangle \Big|_{\vec{y}=\vec{x}} = 2 \sum_{|n\rangle} \xi_{|n\rangle} \langle 0 | \hat{\rho}(\vec{x}) | n \rangle \langle n | \hat{\rho}(\vec{x}) | 0 \rangle = 2 \sum_{|n\rangle} \xi_{|n\rangle} \left| \langle 0 | \hat{\rho}(\vec{x}) | n \rangle \right|^2$$

$$\tag{4.19}$$

Now, in general, the quantity $\langle 0|\hat{\rho}(\vec{x})|n\rangle$ is not zero [11] and since $\xi_{|n\rangle} > 0$ for all

$|n\rangle \neq |0\rangle$ the above expression is non-zero and positive. Therefore $\langle 0|\left[\hat{\vec{J}}(\vec{x}), \hat{\rho}(\vec{y})\right]|0\rangle$ is

non-zero. The key reason for this result is the relationship given by (2.30).

When this result is used in (4.12) is evident that $\delta_g \vec{J}_{vac}(\vec{x}, t)$ is not zero for an arbitrary $\chi(\vec{y}, t)$. Therefore the vacuum current is not gauge invariant and a calculation of the vacuum current should yield non-gauge invariant terms. As explained in the introduction this is always the case.

## V. Redefining the vacuum state

As can be seen from the previous discussion the fact that the vacuum state $|0\rangle$ is a lower bound to the free field energy means that Dirac field theory is not gauge invariant as formulated. As a result of this non-gauge invariant terms appear in the results of calculations of certain quantities such as the vacuum current. These terms must be removed to obtain a physically correct result.

The question that will address in this section is whether it is possible to formulate

Dirac field theory so that the theory is gauge invariant from the start. It has been shown in [9] that this can be done by redefining the vacuum state in such a way that it is no longer a lower bound to the free field energy.

Consider, first, the vacuum state $|0\rangle$ as defined in Section II by equation (2.17).

The top edge of the negative energy band has an energy of $-m$ where m is the mass of

the electron. For the vacuum state $|0\rangle$ all states with an energy less than or equal to $-m$

are occupied and all positive energy states are unoccupied. This definition will be modified as follows: Define the "modified" vacuum state $\left|0_c\right\rangle$ as the quantum state in which each negative energy state in the band from $-m$ to an energy of $-E_c$ is occupied and negative energy states with energy less than $-E_c$ are unoccupied, where $E_c \to \infty$. Also all positive energy states are unoccupied. The state $\left|0_c\right\rangle$ is defined by,

$$\left|0_c\right\rangle = \prod_{n \in \text{band}} \hat{a}_n^\dagger \left|0, \text{bare}\right\rangle \tag{5.1}$$

where the notation $n \in \text{band}$ means the product is taken over all states in the band of states whose energy is between $-m$ and $-E_c$. This can also be expressed as,

$$\begin{aligned} \hat{a}_n \left|0_c\right\rangle &= 0 \text{ for } \lambda_n = 1 \\ \hat{a}_n^\dagger \left|0_c\right\rangle &= 0 \text{ for } -m \geq \lambda_n E_{\vec{p}_n} \geq -E_c \\ \hat{a}_n \left|0_c\right\rangle &= 0 \text{ for } -E_c > \lambda_n E_{\vec{p}_n} \end{aligned} \tag{5.2}$$

Note the state $\left|0_c\right\rangle$ is almost identical to $\left|0\right\rangle$ with the exception that for $\left|0_c\right\rangle$ the bottom of the negative energy band is handled by a limiting process. The cutoff energy $-E_c$ is assumed to finite and is taken to negative infinity at the end of a calculation. States with less energy then $-E_c$ are unoccupied. If one of these states becomes occupied then the new state will have less free field energy then $\left|0_c\right\rangle$. Therefore $\left|0_c\right\rangle$ is no longer the lower bound to the free field energy. In fact there is no lower bound to the free field energy. As shown in Section III this is a necessary requirement for Dirac field theory to be gauge invariant.

If $\left|0_c\right\rangle$ is used as the vacuum state instead of $\left|0\right\rangle$ then the change in the vacuum current is obtained by replacing $\left\langle 0\right|$ and $\left|0\right\rangle$ with $\left\langle 0_c\right|$ and $\left|0_c\right\rangle$, respectively, in (4.1).

The change in the vacuum current due to a gauge transformation is obtained by making the same substitution in (4.12) to obtain,

$$\delta_g \vec{J}_{c,vac}(\vec{x},t) = i \int \langle 0_c | \left[ \hat{\vec{J}}(\vec{x}), \hat{\rho}(\vec{y}) \right] | 0_c \rangle \chi(\vec{y},t) d\vec{y} \tag{5.3}$$

In order for the above quantity to be zero for arbitrary $\chi$ then $\langle 0_c | \left[ \hat{\vec{J}}(\vec{x}), \hat{\rho}(\vec{y}) \right] | 0_c \rangle$ must equal zero. To evaluate this quantity use (2.8) to obtain,

$$\langle 0_c | \left[ \hat{\vec{J}}(\vec{x}), \hat{\rho}(\vec{y}) \right] | 0_c \rangle = \left( \frac{q}{2} \right)^2 \langle 0_c | \left[ \left[ \hat{\psi}^\dagger(\vec{x}), \vec{\alpha}\hat{\psi}(\vec{x}) \right], \left[ \hat{\psi}^\dagger(\vec{y}), \hat{\psi}(\vec{y}) \right] \right] | 0_c \rangle \tag{5.4}$$

Refer to (2.9) to yield,

$$\langle 0_c | \left[ \hat{\vec{J}}(\vec{x}), \hat{\rho}(\vec{y}) \right] | 0_c \rangle = \frac{q^2}{4} \sum_{nmrs} \langle 0_c | \left[ \left[ \hat{a}_r^\dagger, \hat{a}_s \right], \left[ \hat{a}_n^\dagger, \hat{a}_m \right] \right] | 0_c \rangle \left( \phi_r^\dagger(\vec{x}) \vec{\alpha} \phi_s(\vec{x}) \right) \left( \phi_n^\dagger(\vec{y}) \phi_m(\vec{y}) \right) \tag{5.5}$$

Use the anticommutator relationships (2.10) to obtain,

$$\left[ \left[ \hat{a}_r^\dagger, \hat{a}_s \right], \left[ \hat{a}_n^\dagger, \hat{a}_m \right] \right] = 2 \left( \left( \hat{a}_r^\dagger \hat{a}_m - \hat{a}_m \hat{a}_r^\dagger \right) \delta_{sn} + \left( \hat{a}_s \hat{a}_n^\dagger - \hat{a}_n^\dagger \hat{a}_s \right) \delta_{rm} \right) \tag{5.6}$$

Use $\hat{a}_m \hat{a}_r^\dagger = \delta_{rm} - \hat{a}_r^\dagger \hat{a}_m$ in the above to obtain,

$$\left[ \left[ \hat{a}_r^\dagger, \hat{a}_s \right], \left[ \hat{a}_n^\dagger, \hat{a}_m \right] \right] = 2 \left( \left( 2\hat{a}_r^\dagger \hat{a}_m - \delta_{rm} \right) \delta_{sn} + \left( \delta_{sn} - 2\hat{a}_n^\dagger \hat{a}_s \right) \delta_{rm} \right) \tag{5.7}$$

This yields,

$$\left[ \left[ \hat{a}_r^\dagger, \hat{a}_s \right], \left[ \hat{a}_n^\dagger, \hat{a}_m \right] \right] = 4 \left( \hat{a}_r^\dagger \hat{a}_m \delta_{sn} - \hat{a}_n^\dagger \hat{a}_s \delta_{rm} \right) \tag{5.8}$$

Use this in (5.5) to obtain,

$$\left\langle 0_c \left| \left[ \hat{\bar{J}}(\vec{x}), \hat{\rho}(\vec{y}) \right] \right| 0_c \right\rangle = q^2 \sum_{nmrs} \left\langle 0_c \left| \begin{pmatrix} \delta_{ns} \hat{a}_r^\dagger \hat{a}_m \\ -\delta_{mr} \hat{a}_n^\dagger \hat{a}_s \end{pmatrix} \right| 0_c \right\rangle \left( \phi_n^\dagger(\vec{y}) \phi_m(\vec{y}) \right) \left( \phi_r^\dagger(\vec{x}) \vec{\alpha} \phi_s(\vec{x}) \right)$$

$$(5.9)$$

Use (5.2) and (2.10) in the above and redefine some of the dummy variables to yield,

$$\left\langle 0_c \left| \left[ \hat{\bar{J}}(\vec{x}), \hat{\rho}(\vec{y}) \right] \right| 0_c \right\rangle = q^2 \sum_{s \in band} \sum_m \left\{ \begin{array}{l} \left( \phi_s^\dagger(\vec{x}) \vec{\alpha} \phi_m(\vec{x}) \right) \left( \phi_m^\dagger(\vec{y}) \phi_s(\vec{y}) \right) \\ -\left( \phi_s^\dagger(\vec{y}) \phi_m(\vec{y}) \right) \left( \phi_m^\dagger(\vec{x}) \vec{\alpha} \phi_s(\vec{x}) \right) \end{array} \right\}$$

$$(5.10)$$

The notation $s \in band$ means the index 's' is summed over the states whose energy is in the band from $-m$ to $-E_c$. Note that the summation over 'm' is over all states.

Take the summation over 'm' and use (2.15) in the above to obtain,

$$\left\langle 0_c \left| \left[ \hat{\bar{J}}(\vec{x}), \hat{\rho}(\vec{y}) \right] \right| 0_c \right\rangle = q^2 \sum_{s \in band} \left\{ \begin{array}{l} \left( \phi_s^\dagger(\vec{x}) \vec{\alpha} \phi_s(\vec{y}) \right) \delta^{(3)}(\vec{x} - \vec{y}) \\ -\left( \phi_s^\dagger(\vec{y}) \vec{\alpha} \phi_s(\vec{x}) \right) \delta^{(3)}(\vec{x} - \vec{y}) \end{array} \right\}$$

$$(5.11)$$

Next use the relationship,

$$f(\vec{y}) \delta^{(3)}(\vec{x} - \vec{y}) = f(\vec{x}) \qquad (5.12)$$

to obtain,

$$\left\langle 0_c \left| \left[ \hat{\bar{J}}(\vec{x}), \hat{\rho}(\vec{y}) \right] \right| 0_c \right\rangle = q^2 \sum_{s \in band} \delta^{(3)}(\vec{x} - \vec{y}) \left\{ \begin{array}{l} \left( \phi_s^\dagger(\vec{x}) \vec{\alpha} \phi_s(\vec{x}) \right) \\ -\left( \phi_s^\dagger(\vec{x}) \vec{\alpha} \phi_s(\vec{x}) \right) \end{array} \right\} = 0 \qquad (5.13)$$

When this result is used in (5.3) it is seen that $\delta_g \vec{J}_{c,vac}(\vec{x}, t)$ is zero. This means that the vacuum current is gauge invariant when $|0_c\rangle$ is used as the vacuum state.

## VI. Summary and Conclusion

Some problems associated with gauge invariance in Dirac field theory have been discussed. As was stated in the Introduction it is well know that calculations in Dirac field theory can produce non-gauge invariant results. These results cannot be physically valid because a physical theory must be gauge invariant. In order to produce a physically acceptable answer the results of these calculations must be "corrected" by removing the non-gauge invariant parts.

The issue that is raised in this paper is that in the physics textbooks there is no really adequate explanation as to why this correction is required. Why does the theory not produce a gauge invariant result directly if the underlying theory is gauge invariant? This question is addressed in Section III where it is shown that for Dirac field theory to be gauge invariant there must be no lower bound to the free field energy. However, in standard field theory, there is a lower bound to the free field energy which is the energy of the vacuum state. Therefore the underlying theory is *not* gauge invariant. Since the underlying theory is not gauge invariant the results of calculations should have non-gauge invariant terms which, as was discussed in the Introduction, is the case.

It is then shown in Section IV the impact that this has on calculations of the vacuum current. In order for the vacuum current to be gauge invariant the quantity $\left\langle 0 \left| \left[ \hat{\vec{J}}(\vec{x}), \hat{\rho}(\vec{y}) \right] \right| 0 \right\rangle$ must be zero. But as shown, this quantity cannot be zero, due to the fact that the vacuum state $\left| 0 \right\rangle$ is a lower bound to the free field energy.

A possible way to produce a gauge invariant theory is to modify the definition of the vacuum state so that there is no longer a lower bound to the free field energy. This is discussed in Section V where the modified vacuum state $\left| 0_c \right\rangle$ is defined. The state $\left| 0_c \right\rangle$ is similar to the standard definition $|0\rangle$ except that the bottom of the negative energy

band is defined using a limiting procedure for $|0_c\rangle$. In this case the quantity

$\langle 0_c |\left[\hat{\vec{J}}(\vec{x}), \hat{\rho}(\vec{y})\right]| 0_c\rangle$ is zero so that the vacuum current will be gauge invariant.

**Appendix A**

In order to use perturbation theory we must convert from the Schrödinger picture to the interaction picture (see Chapter 4-2 of [12]). Write the Hamiltonian as,

$$\hat{H} = \hat{H}_0 + \hat{V} \tag{A.1}$$

where $\hat{H}_0$ is the unperturbed free field Hamiltonian and $\hat{V}$ is the perturbation. From

(2.6),

$$\hat{V} = -\int \hat{J}(\vec{x}) \cdot \vec{A}(\vec{x}, t) d\vec{x} + \int \rho(\vec{x}) \cdot A_0(\vec{x}, t) d\vec{x} \tag{A.2}$$

Define the interaction state vector by,

$$|\Omega_I\rangle = e^{i\hat{H}_0 t} |\Omega\rangle \tag{A.3}$$

Interaction operators $\hat{O}_I$ are defined in terms of Schrödinger operators $\hat{O}$ according to,

$$\hat{O}_I = e^{i\hat{H}_0 t} \hat{O} e^{-i\hat{H}_0 t} \tag{A.4}$$

The expectation values of operators have the same value in both representations, i.e.,

$O_e = \langle \Omega_I | \hat{O}_I | \Omega_I \rangle = \langle \Omega | \hat{O} | \Omega \rangle$. Use the above expressions in (2.1) to obtain,

$$i\frac{\partial |\Omega_I\rangle}{\partial t} = \hat{V}_I |\Omega_I\rangle \tag{A.5}$$

A formal solution of above is given by,

$$|\Omega_I(t)\rangle = \left(1 - i\int_{t_0}^t \hat{V}_I(t_1) dt_1 + (-i)^2 \int_{t_0}^t \hat{V}_I(t_1) dt_1 \int_{t_0}^{t_1} \hat{V}_I(t_2) dt_2\right) |\Omega_I(t_0)\rangle \tag{A.6}$$

The current expectation value is then given by,

$$\vec{J}_e\left(\vec{x},t\right) = \left\langle \Omega_I\left(t\right) \middle| \hat{\vec{J}}_I\left(\vec{x},t\right) \middle| \Omega_I\left(t\right) \right\rangle$$

$$= \left\langle \Omega_I\left(t_0\right) \middle| \hat{\vec{J}}_I\left(\vec{x},t\right) \middle| \Omega_I\left(t_0\right) \right\rangle - i\left\langle \Omega_I\left(t_0\right) \middle| \left[ \hat{\vec{J}}_I\left(\vec{x},t\right), \int_{t_0}^{t} \hat{V}_I\left(t_1\right) dt_1 \right] \middle| \Omega_I\left(t_0\right) \right\rangle + O\left(\hat{V}^2\right)$$

$$(A.7)$$

where $O\left(\hat{V}^2\right)$ refers to terms to the second order of the perturbation or higher. Let the initial state, at time $t = t_0$, be the vacuum state $|0\rangle$. The first order vacuum current at time t is then,

$$\vec{J}_{vac}\left(\vec{x},t\right) \simeq \left\langle 0_I \middle| \hat{\vec{J}}_I\left(\vec{x},t\right) \middle| 0_I \right\rangle - i\left\langle 0_I \middle| \left[ \hat{\vec{J}}_I\left(\vec{x},t\right), \int_{t_0}^{t} \hat{V}_I\left(t_1\right) dt_1 \right] \middle| 0_I \right\rangle \qquad (A.8)$$

where $|0_I\rangle$ is the interaction vacuum state and is given by,

$$|0_I\rangle = e^{i\hat{H}_0 t}|0\rangle = |0\rangle \qquad (A.9)$$

It is easy to show that $\left\langle 0_I \middle| \hat{\vec{J}}_I\left(\vec{x},t\right) \middle| 0_I \right\rangle = 0$. Use this, along with (A.9), and let the initial time $t_0 = -\infty$ to obtain equation (4.1) in the text.

## **Appendix B**

We will prove that,

$$\frac{\partial \hat{\rho}_I\left(\vec{x},t\right)}{\partial t} = -\vec{\nabla} \cdot \hat{\vec{J}}_I\left(\vec{x},t\right) \qquad (B.1)$$

Use (4.2) in the above to obtain,

$$i\left[\hat{H}_0, \hat{\rho}\left(\vec{x}\right)\right] = -\vec{\nabla} \cdot \hat{\vec{J}}\left(\vec{x}\right) \qquad (B.2)$$

From (2.26) and (2.21) we obtain the following relationships,

$$\left[\hat{H}_0, \hat{b}_j\right] = -E_j\hat{b}_j; \quad \left[\hat{H}_0, \hat{b}_j^\dagger\right] = E_j\hat{b}_j^\dagger; \quad \left[\hat{H}_0, \hat{d}_j\right] = -E_j\hat{d}_j; \quad \left[\hat{H}_0, \hat{d}_j^\dagger\right] = E_j\hat{d}_j^\dagger \qquad (B.3)$$

Use this in (2.19) and (2.20) to obtain,

$$\left[\hat{H}_0, \hat{\psi}(\vec{x})\right] = -H_0\hat{\psi}(\vec{x}); \quad \left[\hat{H}_0, \hat{\psi}^\dagger(\vec{x})\right] = \left(H_0\hat{\psi}(\vec{x})\right)^\dagger \tag{B.4}$$

From this and (2.8) we obtain,

$$\left[\hat{H}_0, \hat{\rho}\right] = \frac{q}{2}\left\{\left[\left(H_0\hat{\psi}\right)^\dagger, \hat{\psi}\right] - \left[\hat{\psi}^\dagger, \left(H_0\hat{\psi}\right)\right]\right\} \tag{B.5}$$

Use (2.5) in the above to obtain,

$$\left[\hat{H}_0, \hat{\rho}\right] = \frac{q}{2}\left\{\begin{bmatrix}\left(-i\vec{\alpha}\cdot\vec{\nabla}\hat{\psi} + m\beta\hat{\psi}\right)^\dagger, \hat{\psi}\end{bmatrix} \\ -\left[\hat{\psi}^\dagger, \left(-i\vec{\alpha}\cdot\vec{\nabla}\hat{\psi} + m\beta\hat{\psi}\right)\right]\right\} \tag{B.6}$$

This becomes,

$$\left[\hat{H}_0, \hat{\rho}\right] = \frac{q}{2}\left\{\begin{bmatrix}\left(-i\vec{\alpha}\cdot\vec{\nabla}\hat{\psi}\right)^\dagger, \hat{\psi}\end{bmatrix} \\ -\left[\hat{\psi}^\dagger, \left(-i\vec{\alpha}\cdot\vec{\nabla}\hat{\psi}\right)\right]\right\} = i\frac{q}{2}\vec{\nabla}\cdot\left[\hat{\psi}^\dagger, \vec{\alpha}\hat{\psi}\right] \tag{B.7}$$

Use (2.8) in the above to obtain (B.2). Therefore (B.1) is also valid.